# PGRF：Physics-Guided Rectified Flow for Low-light RAW Image Enhancement

Juntai Zeng [1]

[1] School of Physics, Northeast Normal University, Changchun 130024, China; jtzeng68@163.com

**Abstract**

Enhancing RAW images captured under low-light conditions is a challenging task. Recent deep learning–based RAW enhancement methods have shifted from using real paired data to relying on synthetic datasets. These synthetic datasets are typically generated by physically modeling sensor noise, but existing approaches often consider only additive noise, ignore multiplicative components, and rely on global calibration that overlooks pixel-level manufacturing variations. As a result, such methods struggle to accurately reproduce real sensor noise. To address these limitations, this paper derives a noise model from the physical noise generation mechanisms that occur under low illumination and proposes a novel composite model that integrates both additive and multiplicative noise. To solve the model, we introduce a physics-based per-pixel noise simulation and calibration scheme that estimates and synthesizes noise for each individual pixel, thereby overcoming the restrictions of traditional global calibration and capturing spatial noise variations induced by microscopic CMOS manufacturing differences. Motivated by the strong performance of rectified-flow methods in image generation and processing, we further combine the physics-based noise synthesis with a rectified-flow generative framework and present PGRF (Physics-Guided Rectified Flow): a physics-guided rectified-flow framework for low-light image enhancement. PGRF leverages the ability of rectified flows to model complex data distributions and uses physical guidance to steer the generation toward the desired clean image. To validate the effectiveness of the proposed model, we established the LLID dataset, an indoor low-light benchmark captured with the Sony A7S II camera. Experimental results demonstrate that the proposed framework achieves significant improvements in low-light RAW image enhancement.

## 1. Introduction

In recent years, driven by growing demands for higher image quality, low-light image enhancement has attracted widespread attention. Deep learning–based enhancement methods have achieved great success and gradually become the dominant approach [1–3]. These methods typically learn a mapping from low-light images to their corresponding long-exposure (clean) images, and they perform well in detail restoration and noise suppression. Although most approaches operate in the sRGB domain and achieve reasonable results, the sRGB space compresses sensor-native information and therefore cannot fully exploit the potential of the underlying image data [4]. By contrast, the RAW domain has been gaining increasing interest because it preserves higher bit depth and the sensor's original noise characteristics more directly [5,6]. Nevertheless, restoring RAW images captured under low illumination remains highly challenging. A primary reason is that learning-based methods require large amounts of paired noisy–clean training data; collecting such datasets is laborious because it is necessary to ensure perfect alignment between the noisy and clean images, with no spatial misregistration. To address this problem, synthesizing paired datasets by physically modeling the camera's noise generation process and using those synthetic pairs to train networks has emerged as an efficient solution [7–9].

In low-light conditions, physics-based image sensor noise modeling seeks to represent noise by accounting for the physical generation mechanisms and the statistical properties of the sensor. Currently, physics-based modeling methods, exemplified by ELD [7], typically consider only additive noise while neglecting multiplicative noise—that is, the multiplicative relationship between dark current shot noise and fixed pattern noise. Moreover, these methods mainly model noise with respect to frame-to-frame relationships, using a single noise value to generate a global noise pattern for the image sensor (hereafter referred to as global calibration). They do not take into account that, as pixel sizes in image sensors become increasingly small, it becomes more difficult to manufacture identical pixels [10], meaning that the noise generated by each individual pixel in the image sensor may vary. Consequently, the modeling of image sensor noise may lack precision. To address these shortcomings, we formulate a composite noise model that incorporates both additive and multiplicative components. Furthermore, we abandon global calibration in favor of a novel per-pixel calibration strategy: by observing the noise behavior of individual pixels and fitting our composite model at the pixel level, we estimate and synthesize pixel-specific noise patterns. This per-pixel noise simulation more accurately reproduces the sensor's actual noise and yields higher-quality paired training samples.

Meanwhile, generative models, particularly diffusion-based approaches [11,12], have demonstrated significant advantages in image enhancement by leveraging their powerful generation capabilities to produce high-quality image details. However, their cumbersome iterative process results in slow inference, which severely limits their industrial applicability. Recently, the rectified flow generative model, Rectified Flow [13], has demonstrated strong performance, capable of rapidly generating images with minimal loss of detail. Nevertheless, conventional rectified flow models exhibit a one-to-many mapping, making them unsuitable for low-light image enhancement, which is inherently a one-to-one mapping task. To address this issue, we propose a physics-guided rectified flow framework for low-light image enhancement—PGRF (Physics-Guided Rectified Flow). To mitigate the randomness inherent in Rectified Flow during generation, and inspired by [11], we design a physics-guided conditional control module that steers the model toward generating a specific target image. To further enhance the generative capability of our framework, we introduce a sampling search strategy after analyzing the rectification strategy of Rectified Flow, which strengthens the model's ability to accurately model the generation path. By integrating physical priors with the conditional control mechanism, we effectively address the adaptability of rectified flow models in one-to-one mapping tasks while improving generation quality.

The proposed framework is simple and efficient. Compared with diffusion-based low-light image enhancement methods, PGRF does not require multiple iterations, thereby avoiding long inference times. Moreover, the rectified flow-based approach can efficiently estimate the mapping path from a normal distribution to the target image, enabling rapid generation and achieving a better balance between efficiency and performance. To the best of our knowledge, this is the first work to combine physics-guided paired data generation with rectified flow models, thereby avoiding the difficulty of collecting paired data while leveraging the powerful generative capability to achieve superior low-light image restoration.

Main contributions:
- Pixel-level sensor noise model with multiplicative component.

We develop a noise model targeted at image-sensor pixels that extends traditional formulations by incorporating multiplicative noise. This model more accurately reflects the physical noise generation mechanisms of image sensors and provides a practical approach for parameter estimation and model solution.

- Per-pixel calibration and novel noise-simulation strategy.

To accommodate the proposed physical noise model, we shift the noise-estimation perspective from conventional global calibration to per-pixel calibration, introducing a novel noise-synthesis scheme. This scheme operates from the spatial dimension, combining a physical model to individually estimate the noise of each pixel in the sensor, and generates more realistic noise for each pixel during the noise simulation phase, thereby improving simulation accuracy.

- Physics-Guided Rectified Flow Framework for Low-Light Enhancement (PGRF).

We further integrate physics-based modeling with the rectified flow generative model, proposing a physics-guided rectified flow framework for low-light image enhancement (PGRF). This method fully exploits the strengths of generative models in image generation, resulting in enhanced low-light images that are more natural and realistic.

- A new indoor low-light dataset for evaluation.

We collected and constructed an indoor low-light image dataset, LLID, covering various shooting settings and scenes. This dataset is used for evaluating enhancement performance and comparing the effectiveness of different methods, providing a solid data foundation for this study.

## 2. Recent Work

### 2.1 RAW Low-Light Image Processing

RAW-domain low-light image enhancement has received considerable attention in recent years [3,4,6,14]. Chen et al. [3] drew wide interest to this field by designing a dedicated U-Net architecture and collecting the SID dataset. To address the difficulty of acquiring paired RAW data, synthesizing paired data has since become a mainstream research direction [7,15–22]. Foi et al. [23] pioneered a Poisson–Gaussian noise model suitable for raw data, modeling image noise as a mixture of Poisson and Gaussian components; however, this model faces severe challenges under low-illumination imaging conditions [24]. Wei et al. [7], by analyzing the imaging mechanisms of camera noise in low-light environments, proposed a physics-based noise model that synthesizes the noise distribution of low-light images to generate paired data; their approach achieved results that even surpass models trained on real paired datasets and has drawn further attention to synthetic paired-data methods.

However, noise modeling based purely on physical methods mostly considers only additive noise, neglecting the nonlinear relationship between dark current shot noise and fixed pattern noise, and thereby ignoring multiplicative noise. The principal reason is the limitation of existing calibration techniques: estimating multiplicative noise is difficult in practice. To address this shortcoming, we adopt a pure-physics modeling perspective to formulate an additive–multiplicative noise model and combine it with a novel calibration method to estimate the multiplicative noise term.

Abdelhamed et al. [15] introduced a noise-generation framework based on conditional normalizing flows (Noise Flow), which for the first time combined a physics-driven parametric noise model with a deep generative model. Their approach constructs an invertible flow network composed of a signal-dependent layer (to model heteroscedastic noise) and a gain layer (to model the sensor ISO response), enabling joint modeling of multi-device, multi-ISO noise on the SIDD dataset. Monakhova et al. [16] proposed a novel noise modeling method that combines Generative Adversarial Networks (GANs) with physical modeling. The method dynamically adjusts the noise parameters through the GAN based on the types of noise produced by the image sensor, and constrains the generated noise using a discriminator. Zhang et al. [18] combined physical modeling with GANs and proposed a general noise model with a decoupled synthesis process. The model generates noise using the generative network and employs a Fourier transform-based discriminator to determine whether the noise conforms to the characteristics of sensor noise. Qin et al. [21] employed diffusion models for noise modeling, generating noise by progressively adding noise in multiple steps and then reversing this process via denoising. Li et al. [25] proposed a hypothesis-based shot noise synthesis method, which can generate noise using only captured dark frames.

Among them, learning-based methods mostly train neural networks on real paired data to construct noise surrogate models. However, their modeling performance is limited due to the inaccuracy of distribution measurement mechanisms [22]. Moreover, most learning-based methods rely on paired real datasets for training, which conflicts with our goal of avoiding the difficulties of dataset collection. Therefore, we focus primarily on physics-based sensor noise modeling, which does not depend on paired data but estimates sensor noise through physical methods.

*2.2 Low-Light Image Enhancement Based on Generative Models*

In recent years, low-light image enhancement methods based on GANs have attracted considerable attention [26–28]. However, GAN-based approaches suffer from inherent instability and the need for careful parameter tuning, which pose significant challenges. With the introduction of diffusion models [29], their strong generative capacity and improved stability have gradually established them as a mainstream framework. Owing to their ability to generate high-quality image details, researchers have begun to explore the application of diffusion models to low-light image enhancement [4, 11, 12, 30].

Zhou et al. [11] addressed the challenges of complex sampling procedures and long processing times in diffusion models by proposing a pyramid diffusion model, which achieved notable improvements in both sampling efficiency and restoration performance. Hou et al. [12] introduced a curvature regularization term to constrain the curvature of the corresponding ODE in the diffusion process, maintaining a low-curvature trajectory to prevent excessive divergence and instability during generation. Furthermore, they incorporated non-local structural information from the image data to construct a global structure-aware regularization term, which preserved overall structural consistency while effectively enhancing image contrast and strengthening detail representation. Jiang et al. [30] combined diffusion models with wavelet transforms and proposed a wavelet-based conditional diffusion model. Specifically, they decomposed the input image via multi-level two-dimensional discrete wavelet transform, thereby restricting the diffusion process to the wavelet domain. This approach compressed the spatial resolution and reduced computational complexity, significantly accelerating the inference speed of the diffusion model. Li et al. [4] explored diffusion models for RAW-domain low-light image enhancement and proposed a two-stage framework consisting of pre-training and alignment. In the pre-training stage, multiple virtual camera noise models were constructed in the noise space, and a Camera Feature Integration (CFI) module was introduced to learn cross-camera general feature representations. In the alignment stage, a small number of real paired RAW images were used for fine-tuning to adapt the model to device-specific noise characteristics. To further address the common issue of color shifts in diffusion processes, a color corrector was introduced to dynamically adjust the global color distribution, thereby improving the color consistency of the final enhanced image.

## 3. Mathematical Modeling of Image Sensors

Our analysis focuses primarily on CMOS image sensors. First, we examine the physical mechanisms through which individual sensor pixels generate noise. Based on this analysis, we construct a mathematical model that incorporates both additive and multiplicative relationships among different noise components, thereby more accurately reflecting the noise generated at each pixel. The proposed noise model specifically targets low-light imaging conditions and assumes that the exposure time during image acquisition is longer than the minimum exposure time of the sensor, under which characteristic noise is produced.

*3.1 Noise Analysis*

3.1.1 Photon Shot Noise

Photon shot noise, also referred to as quantum noise, primarily arises from the discrete and statistical nature of photons. Since photons behave as particles with quantum properties, their arrival at the image sensor follows a random process. This randomness causes the number of incident photons to fluctuate around an expected mean value rather than remaining constant, thereby introducing noise. Photon shot noise is widely recognized to follow a Poisson distribution [31]. Its mathematical expression is given by:

$$(I(x,y) + N_s(x,y)) \sim P(\lambda) \tag{3.1}$$

$I(x,y)$ is the true photon count of the target scene. $N_s(x,y)$ denotes the photon shot noise. This noise is inherent to the fundamental nature of light and is unavoidable.

3.1.2 Read Noise

Read noise is defined as all noise components that do not vary with the incident light signal [32]. It includes dark current shot noise, fixed-pattern noise, row noise, source follower noise, and quantization noise.

3.1.3 Dark Current Shot Noise

Dark current mainly arises from thermally generated electron–hole pairs. Even in the absence of illumination, these electrons and holes are randomly generated and move within the sensor, thereby forming dark current. Its physical expression is given in [33] as:

$$S = QT^{3/2} C \exp\left(-\frac{E_{gap}}{2 \cdot kT}\right) \tag{3.2}$$

Where $Q$ denotes the pixel area, $T$ represents the absolute temperature, and $C$ is the dark current figure-of-merit at 300 K, which varies across different sensors and is usually provided by the manufacturer. $E_{gap}$ denotes the band gap energy of the semiconductor, and $k$ is the Boltzmann constant.

This type of noise induced by dark current is referred to as dark current shot noise. It originates from defects introduced during the semiconductor fabrication process. Consequently, the magnitude of dark current varies across different pixels, and it also depends on exposure time—specifically, under fixed conditions for other parameters, a longer exposure time results in stronger dark current noise. Such noise is generally modeled by a Poisson distribution [34]:

$$N_{RS}(x,y) = t \cdot S \tag{3.3}$$

Its expression is:

$$N_{RS}(x,y) \sim P(\lambda) \tag{3.4}$$

$N_{RS}(x,y)$ represents the dark current shot noise.

3.1.4 Fixed-Pattern Noise

Fixed-pattern noise (FPN) primarily arises from manufacturing process variations in image sensors, such as differences in transistor threshold voltages and non-uniform doping concentrations, which lead to pixel-dependent deviations. In captured images, FPN manifests as noise points fixed at specific pixel locations. Unlike random noise components, FPN remains stable across consecutive frames and therefore cannot be effectively suppressed by temporal averaging. A common strategy for mitigating FPN is to subtract the bias estimated from dark-frame captures. Under low-light conditions, however, FPN becomes more pronounced and exerts a stronger detrimental effect on image quality. From the perspective of the entire sensor, FPN is generally modeled as following a Gaussian distribution, expressed as:

$$N_{FP}(x,y) \sim N(0,\sigma^2) \tag{3.5}$$

$N_{FP}(x,y)$ represents the fixed-pattern noise (FPN).

Since dark current shot noise follows a Poisson distribution, we assume it to be multiplicative noise. Inspired by [24], we use the following relation to describe the relationship between dark current shot noise and fixed-pattern noise.

$$N_d = N_{RS}(x,y) + N_{RS}(x,y) \cdot N_{FP}(x,y) \tag{3.6}$$

Where $N_d$ a denotes the total contribution of dark current shot noise and fixed-pattern noise, $N_{RS}(x,y)$ represents the dark current shot noise, which follows a Poisson distribution, and $N_{FP}(x,y)$ denotes the fixed-pattern noise, which is generally modeled as a Gaussian distribution.

3.1.5 Row Noise

In CMOS sensors, each row is typically controlled by a dedicated analog-to-digital converter (ADC) during signal readout. Slight fluctuations and instabilities in the operating voltages of different ADCs lead to inconsistencies in signal intensity, thereby giving rise to row noise. Row noise is generally assumed to follow a Gaussian distribution [35], which can be expressed as:

$$N_H(x,y) \sim N(0,\sigma^2) \tag{3.7}$$

$N_H(x,y)$ represents the row noise distribution.

3.1.6 Source Follower Noise

Source follower noise primarily arises from defects associated with the random capture and release of charge carriers in the silicon lattice. It mainly consists of thermal noise, flicker noise (1/f noise), and random telegraph noise (RTN). This type of noise is generally modeled as a zero-mean Gaussian distribution [36].

$$N_r(x,y) \sim N(0,\sigma^2) \tag{3.8}$$

$N_r(x,y)$ represents the source follower noise.

3.1.7 Quantization Noise

Quantization noise refers to the error introduced during the analog-to-digital conversion process, arising from the discretization of continuous signals into finite levels. This error is manifested as the difference between the input signal and its quantized representation. Due to the limited precision of the sensor's ADC, signals are quantized before being stored as raw RGB images [17]. Quantization noise is generally modeled by a uniform distribution.

$$N_q(x,y) \sim U(-1/2q, 1/2q,) . \tag{3.9}$$

$N_q(x,y)$ represents the quantization noise, $q$ is the quantization step size, which is typically 1.

*3.2 Model Establishment*

3.2.1 Noise Relationship Analysis

Based on the above analysis of various noise sources and inspired by [7], [24], and [36], we establish the following noise model:

$$D(x, y) = K(I(x,y) + N_s(x,y)) + N_{RS}(x,y)(1 + N_{FP}(x,y)) + N_H(x,y) + N_r(x,y) + N_q(x,y)$$

Here, $(x, y)$ denotes the pixel coordinate; $N_s$ represents photon shot noise; $N_{FP}$ denotes fixed-pattern noise; $N_{RS}$ corresponds to dark current shot noise; $N_H$ denotes row noise; $N_r$ represents source follower noise; and $N_q$ corresponds to quantization noise; K represents the photon conversion rate inside the image sensor.

3.2.2 Calibration Method

In order to solve the various parameters in the model and precisely characterize the noise properties of image sensors. we employ three types of raw images—flat-field frames, bias frames, and dark frames—each designed to capture different sources of noise.

- Flat-field Frame: This image is captured when the sensor is illuminated under uniform lighting conditions. Specifically, the sensor is directed toward a uniformly white surface that is illuminated by a homogeneous light source. To facilitate subsequent computations, the exposure time is set identical to that of the bias frame, and the focal length is adjusted to infinity. The flat-field frame is primarily used to calibrate the uniformity of the sensor's response and provides a reference baseline for subsequent noise analysis.
- Bias Frame: The bias frame refers to an image captured at the shortest exposure time with the sensor lens fully covered, producing a pure black image. The bias frame is primarily used to capture time-independent noise introduced by the sensor circuitry, including fixed-pattern noise and source follower noise.
- Dark Frame: The dark frame is captured under the same exposure time, ISO settings, and temperature conditions, with the sensor lens also covered. Compared to the bias frame, the dark frame not only includes the noise components present in the bias frame but also incorporates dark current shot noise induced by the integration time. Dark frames provide an effective means of characterizing dark current noise under long exposure conditions and are therefore essential for high-precision noise calibration.

In the process of simulating low-light image sensor noise, global calibration is commonly employed to guide the noise simulation workflow. While this approach can reduce the complexity of noise simulation, it exhibits significant limitations in terms of simulation accuracy. During the manufacturing process, image sensors inevitably exhibit process variations, resulting in slight differences in the quantum efficiency and charge collection efficiency of individual pixels, which in turn lead to variations in noise characteristics. Such pixel-level differences are often neglected in global calibration, adversely affecting image quality and fidelity under low-light conditions.

To address this issue, we propose a novel noise-fitting method that performs fine-grained calibration of noise characteristics for each individual pixel and generates noise in a targeted manner. Specifically, the image is calibrated spatially by computing the mean and variance for each pixel at the corresponding position across the image, as shown in Figure 1.

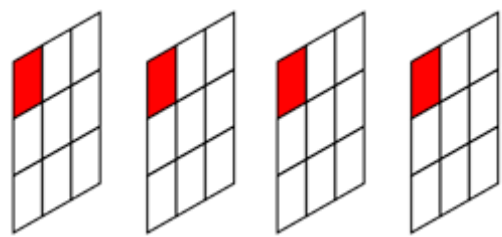

**Figure 1.** Red dots represent the pixel values at corresponding positions across frames.

By comparing single-frame fitting with pixel-wise fitting, we randomly select pixel positions and fit their noise distribution using the Probability Plot Correlation Coefficient (PPCC) [37]. The $R^2$ value reflects the goodness of fit, with larger values indicating better fitting performance. As shown in Figure 2, in most cases, pixel-wise noise fitting outperforms conventional single-frame fitting.

This fine-grained, pixel-level calibration method not only significantly improves the accuracy of noise simulation but also highlights its practical importance by revealing variance differences across pixels. As illustrated in Figure 3, we randomly selected 100 pixels and compared their variances, which showed an amplitude of approximately 3. However, during post-processing with magnification factors ranging from 100 to 200, mathematical analysis revealed that the variance amplification could reach levels between 10,000 and 40,000 times. This further confirms the critical importance of such calibration in low-light imaging.

Compared with traditional approaches that approximate noise using a single variance value, pixel-wise noise generation enables more precise and targeted modeling, thereby providing robust technical support for high-quality imaging under low-light conditions.

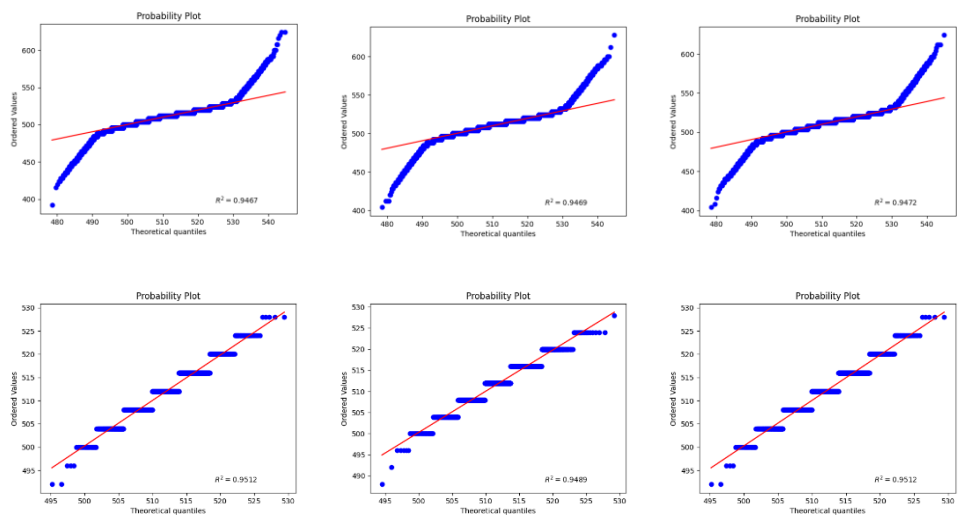

**Figure 2.** Source follower noise variance distribution under ISO 800. Global fitting results are shown in the upper panel, and single-pixel fitting results are shown in the lower panel.

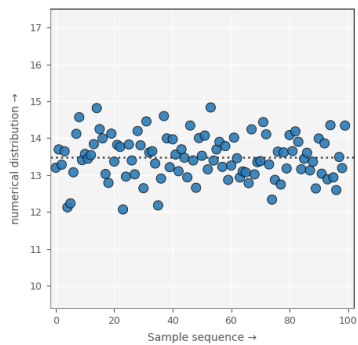

**Figure 3.** Variance distribution of the source follower noise for individual pixels.

Since fixed-pattern noise remains constant across frames, we estimate its value by averaging each pixel over the acquired bias frames. In contrast, for source follower noise, we estimate it by computing the variance across individual pixels within the bias frames. Once the coefficient K and dark current shot noise are estimated through the proposed model, the calibrated data can be employed to generate realistic noise.

3.2.3 Model Solution

We divide the model solution into two parts according to whether the noise components are illumination-dependent. The first part concerns the calibration of photon shot noise, which is primarily estimated using flat-field frames. The second part focuses on the calibration of readout noise, which is derived from dark frames and bias frames. Specifically, the formulation can be expressed as follows:

$$D(x,y) = K\left(I(x,y) + N_s(x,y)\right) + N_{RS}(x,y)\left(1 + N_{FP}(x,y)\right) + N_H(x,y) + N_r(x,y) + N_q(x,y)$$

For simplicity, we write it as:

$$D(x,y) = K(I + N_s) + N \tag{3.10}$$

$$N = N_{RS} \cdot (1 + N_{FP}) + N_H + N_r + N_q \tag{3.11}$$

3.2.4 Photon Shot Noise Calibration

We begin by estimating the unknown variables in Equation (3.10), and subsequently treat Equation (3.11) as an entirety for the estimation of its unknowns. In the first part, our main objective is to estimate the coefficient K Following the approach in [7], we begin by taking the variance on both sides of Eq. (3.10):

$$Var(D) = Var(K(I + N_s)) + Var(N) \tag{3.12}$$

Since $(I + N_s) \sim P(\lambda)$, its variance equals its mean. Therefore:

$$Var(D) = K^2 I + Var(N) \tag{3.13}$$

By simplifying Equation (3.13), we obtain:

$$\frac{Var(D) - Var(N)}{|K|I} = |K| \tag{3.14}$$

Given that $Var(D) \gg Var(N)$, Equation (3.14) can be approximated as:

$$\frac{Var(D)}{|K|I} = |K| \tag{3.15}$$

Here, $Var(D)$ is estimated using the variance of the flat-field frame, while $|K|I$ is estimated using the mean of the flat-field frame. Therefore, all terms on the left-hand side of Eq. (3.15) are known, which allows us to estimate the coefficient K。

3.2.5 Readout Noise Calibration

As discussed in Sections 3.1.3 and 3.1.4, when performing calibration over the entire image sensor, dark current shot noise follows a Poisson distribution, while fixed-pattern noise follows a Gaussian distribution. Consequently, Eq. (3.11) takes the form of a Poisson distribution multiplied by a Gaussian distribution, which makes direct solution intractable. However, since our calibration is performed at the level of individual pixels, we can exploit the fact (as noted in Section 3.1.4) that the fixed-pattern noise of each pixel remains constant. Therefore, it can be treated as a constant term, denoted as $K_2 = (1 + N_{FP})$. By taking the mean of both sides of Eq. (3.11), the expression can be simplified as follows:

$$E(N) = E(N_{RS} \cdot (1 + N_{FP})) + E(N_h) + E(N_r) + E(N_q) \tag{3.16}$$

This can be simplified as:

$$E(N) = K_2 E(N_{RS}) \tag{3.17}$$

Here, $E(N)$ denotes the mean of the dark frame. Since the mean of a Poisson distribution is always non-negative, we take the absolute value of the estimated $N_{RS}$ to obtain the estimated dark current shot noise.

## 4. Enhanced Model Based on Rectified Flow

Given the remarkable performance of rectified flow generative models in the field of image generation and processing, our objective is to exploit their strong generative capability to enhance raw low-light images under the guidance of physical models.

*4.1 Introduction to Rectified Flow*

Rectified Flow [13] is a method that constructs transport mappings between two probability distributions by learning an ordinary differential equation (ODE) trajectory that closely follows a linear interpolation path. This approach is characterized by its simplicity and efficiency, and has been widely applied to tasks such as generative modeling and domain transfer. Specifically, let two probability distributions defined on the Euclidean space $R^d$ be denoted as the source distribution $\pi_0$ and the target distribution $\pi_1$, The transport mapping problem aims to find a deterministic mapping $T: R^d \to R^d$ such that when $X_0 \sim \pi_0$, it holds that $T(X_0) \sim \pi_1$.

Rectified flow realizes the transport mapping from $\pi_0$ to $\pi_1$ by constructing a continuous-time ODE flow along a "straight-line path." Specifically, let $X_0 \sim \pi_0, X_1 \sim \pi_1$, then the linear interpolation path over time $t \in [0,1]$ is defined as:

$$X_t = tX_1 + (1-t)X_0 \tag{4.1}$$

This path corresponds to the shortest trajectory in Euclidean space between $X_0$ and $X_1$. The core idea of rectified flow is to fit a parameterized velocity field $v_\theta(z,t)$, which defines the following form of ordinary differential equation in the continuous-time domain:

$$\frac{dX_t}{dt} = v_\theta(X_t, t) \tag{4.2}$$

Where $X_t$ denotes the trajectory of a random variable evolving with time, and $v_\theta$ parameterized by a neural network, estimates a vector field that best approximates the velocity of the interpolation path. Since $X_t$ is the linear interpolation between $X_0$ and $X_1$, its derivative is constant:

$$\frac{dX_t}{dt} = X_1 - X_0 \tag{4.3}$$

Therefore, rectified flow fits the velocity field $v_\theta$ such that its output along the interpolation path aligns as closely as possible with this derivative direction. Concretely, This differential equation describes the continuous transport process from the source distribution $\pi_0$ to the target distribution $\pi_1$. By differentiating equation (4.1) with respect to t, the target direction $X_1 - X_0$ is obtained, and the velocity field $v_\theta$ can then be fitted by solving the following least-squares problem:

$$\min_\theta \int_0^1 E \left\| (X_1 - X_0) - v_\theta(tX_1 + (1-t)X_0, t) \right\|^2 dt \tag{4.4}$$

This objective can be efficiently optimized via standard stochastic gradient descent, without relying on adversarial training, variational inference, or complex approximate reasoning. As a result, rectified flow theoretically establishes a stable generative model while, in practice, significantly reducing the number of inference steps and improving both the efficiency and quality of image generation and cross-distribution mapping.

*4.2 RAW Image Enhancement Based on Rectified Flow*

Rectified flow models have attracted considerable attention due to their stability and strong generative capability. However, the inherent stochasticity of rectified flow models makes their direct application to deterministic tasks such as image enhancement still challenging. Inspired by [38], we introduce guiding conditions into rectified flow so that it can effectively recover target images. Since the physical noise model in low-light environments inherently contains sensor-specific noise information, our objective is to train the velocity field $v_\theta$ under the guidance of both noisy images generated from this model, thereby enabling accurate recovery of the target image. Consequently, we construct a physics-guided rectified flow framework.

First, we define the base model as:

$$X_t = tX_H + (1-t)X_0 \tag{4.5}$$

$$v_\theta = f(X_t, t) \tag{4.6}$$

Where $X_H$ denotes a high-quality image, $X_0$ a noise sample drawn from a Standard Normal Distribution, and $t \in [0,1]$, $f$ represent processing networks.

To enhance the network's capability for low-light image enhancement, we introduce guiding conditions into the model. Based on the noise distribution characteristics of low-light images, we design a noise model that adds noise to high-quality images, thereby simulating noisy images $X_L$ produced under low-light conditions, such that $X_L = C(X_H)$, where $C(X)$ denotes the noise addition module.

Accordingly, we obtain our guiding condition:

$$T = (X_L) \tag{4.7}$$

We then integrate the guiding condition into the processing network $f$, yielding

$$v_\theta = f(X_t, T, t) \tag{4.8}$$

Subsequently, the trajectory predicted by the velocity field $v_\theta$ is optimized by minimizing the following loss function:

$$L = \min \| X_H - X_0 - v_\theta(X_t, T, t) \|_1 \tag{4.9}$$

*4.3 Sampling Search Strategy*

Inspired by [39], which pointed out that rectified flow models exhibit varying levels of training difficulty across different sampling timesteps, we further hypothesize that, during the sampling phase, the prediction quality may also vary significantly across timesteps. In other words, certain timesteps along the sampling trajectory may play a more critical role in determining the final generation quality. Based on this hypothesis, we propose a sampling step search strategy designed to improve inference performance without modifying the pretrained model parameters.

Specifically, after training is completed, we freeze the model parameters and construct guiding conditions using the high-quality images (the construction method of the guiding conditions is detailed in Section 4.2). During sampling, instead of adopting the random timestep sampling strategy used in training, we introduce a structured equidistant sampling strategy. By controlling the sampling step size $s$ and the total number of steps $n$, the sampling timestep sequence is generated as:

$$t = \{s, 2s, 3s\ldots, ns\}, \ 0 < s < 1, \ 0 < ns < 1$$

Where $s$ and $n$ are tunable parameters that determine the granularity and length of the sampling process.

After performing step-size search and obtaining the optimal sampling time $t_2$, we employ a two-stage sampling mechanism during testing to fully exploit the performance gains from the search. First, we set the time step to $t_1 = 0$, and obtain the intermediate image $X_Z$ based on Equation (4.1) and the trained $v_\theta$.

$$X_Z = v_\theta(X_0, T, t_1) + (1 - t_1)X_0 \tag{4.10}$$

Next, the intermediate result $X_Z$ is combined with the original input $X_0$ through $t_2$ to construct an intermediate state $X_t$:

$$X_t = t_2 X_Z + (1 - t_2) X_0 \tag{4.11}$$

Finally, using the optimal timestep $t_2$ obtained from the search, we invoke the rectified flow prediction process again to generate the final output image $X_M$:

$$X_M = v_\theta(X_t, T, t_2) + X_0 \tag{4.12}$$

The essence of this sampling scheme lies in obtaining a stable intermediate state at timestep $t_1 = 0$, and then leveraging the optimal timestep $t_2$ for a second-stage refinement. This hierarchical two-stage sampling not only preserves the strong generative capability of the rectified flow model but also significantly enhances controllability along the sampling trajectory and improves fine detail reconstruction in the generated images. The results shown in Table 5 validate the effectiveness of our hypothesis. The pseudocode is presented as follows:

| Algorithm 1 Training | |
| --- | --- |
| 1: | Input: target image $X_1$; velocity model $v_\theta$: $R^d \to R^d$ with parameter $\theta$; Physics-Based Noise Model $C(X)$ |
| 2: | Sample: $X_0 \sim N(0,1)$ |
| 3: | Sample: $T = C(X_1)$ |
| 4: | Sample: $t \sim Uniform([0,1])$ |
| 5: | $X_t = tX_1 + (1-t)X_0$ |
| 6: | $\hat{\theta} = \min \|X_1 - X_0 - v_\theta(X_t, T, t)\|_1$ |
| 7: | $dZ_t = v_{\hat{\theta}}(Z_t, T, t) dt$ |
| 8: | Return: $Z = \{Z_t : t \in [0,1]\}$ |

| Algorithm 2 Sampling | |
| --- | --- |
| 1: | Input: target image $X_1$; velocity model $v_\theta$: $R^d \to R^d$ with parameter $\theta$; Physics-Based Noise Model $C(X)$; $t_{best} = 0$; $M_{best} = 0$. |
| 2: | Repeat |
| 3: | Sample: $X_0 \sim N(0,1)$ |
| 4 | Sample: $T = C(X_1)$ |
| 5: | $X_Z = v_\theta(X_0, T, 0) + X_0$ |
| 6: | for $t = s, 2s, 3s…, ns$ do; $0 < s < 1, 0 < ns < 1$ |
| 7: | $X_t = tX_2 + (1-t)X_0$ |
| 8: | $X_M = v_\theta(X_t, T, t) + X_0$ |
| 9: | $M = P(X_M, X_1)$ <br> $P(X_M, X_1)$ means computing the PSNR between $X_M$ and $X_1$ |
| 10: | If $M > M_{best}$; |
| 11: | $M_{best} = M$ |
| 12: | $t_{best} = t$ |
| 13: | Return $t_{best}$ |

| | Algorithm 3 Testing |
|---|---|
| 1: | Input: target image $X_1$; velocity model $v_\theta: R^d \to R^d$ with parameter $\theta$; Physics-Based Noise Model $C(X)$; $t_{best} = 0$; $M_{best} = 0$. |
| 2: | Repeat |
| 3: | Sample: $X_0 \sim N(0,1)$ |
| 4 | Sample: $T = C(X_1)$ |
| 5: | $X_2 = v_\theta(X_0, T, 0) + X_0$ |
| 6: | for $t = s, 2s, 3s\ldots, ns$ do; $0 < s < 1$, $0 < ns < 1$ |
| 7: | $X_t = tX_2 + (1-t)X_0$ |
| 8: | $X_M = v_\theta(X_t, T, t) + X_0$ |
| 9: | $M = P(X_M, X_1)$ <br> $P(X_M, X_1)$ means computing the PSNR between $X_M$ and $X_1$ |
| 10: | If $M > M_{best}$; |
| 11: | $M_{best} = M$ |
| 12: | $t_{best} = t$ |
| 13: | Return $t_{best}$ |

## 5 Experiments

*5.1 Dataset Construction*

For our indoor low-light dataset LLID, we provide a paired noisy–clean dataset captured with a Sony A7S2 camera. To avoid errors caused by camera shake and focus instability, the camera was remotely controlled via Sony's official software, ensuring that no physical contact occurred during the shooting process.

For image acquisition, we followed a strategy similar to that used in ELD [7]: noisy images were captured with high ISO and short exposure times, while the reference clean images were obtained with high ISO and long exposure times. The dataset covers 10 indoor scenes and includes three ISO levels (800, 1600, and 3200) for testing. In total, it contains 160 images.

*5.2 Verification of the Physical Model*

In this section, we provide detailed descriptions of the experimental setup and conduct comprehensive experiments. We compare our method with existing physics-based approaches to demonstrate the superiority of our model and methodology.

5.2.1 Experimental Details

We adopted the same noise simulation strategy as ELD [7], simulating low-light images by reducing clean images by a factor of 200 to 300. Noise was then randomly generated and added to the downscaled images based on the variances of the various noise components estimated using our model.

For the calibration of pixel-level noise, we use the central 1024*1024 region of the captured flat-field, dark-field, and background images. For row noise calibration, since it requires the standard deviation of the mean pixel values across each sensor row, we adopt a global calibration approach when generating row noise. In contrast, fixed-pattern noise, dark current shot noise, source follower noise, and photon shot noise are generated on a per-pixel basis.

When estimating dark current shot noise, in order to avoid frequently capturing dark frames with different exposure times, we chose to capture dark frames with relatively long exposure times to estimate the dark current shot noise. Starting from the generation mechanism of noise electrons within the image sensor, [24] demonstrates that dark current shot noise is proportional to time t. According to Equation (3.2), it can be fitted using $y = at$. However, since the noise generated by the image sensor undergoes complex internal conversions from electrons to voltage and numerical quantization during readout, the corresponding relationship may vary. Therefore, after comprehensive analysis, we propose employing the nonlinear formula $y = a\sqrt{t}$ to fit the shot noise. The comparative results, as presented in Table 1, verify the effectiveness of our fitting approach. Furthermore, to improve robustness across different exposure times under the same ISO setting, we employ random sampling of exposure times during simulation.

This study focuses on calibrating individual pixel points of image sensors, and due to the differences in pixel noise characteristics across different cameras, the network trained using our method may exhibit limited generalization ability. To validate the accuracy of the noise model proposed for low-light noise modeling, we conducted testing using the LLID dataset. To control variables, we adopted the same deep neural network architecture as in [7] for low-light image denoising. [7] utilized a U-Net architecture identical to that in [40] [3], which has been widely used in image denoising tasks. We selected long-exposure images from the SID dataset as the training set, assuming these long-exposure images to be ideal clean images. Based on this assumption, we synthesized corresponding noise on these clean images using the established noise model, thereby generating noise-injected samples for network training. Our implementation is based on PyTorch. We employed a loss function and the Adam optimizer with a batch size of 1, training the model for 200 epochs. The initial learning rate was set to . Our image training and testing pipeline is shown in Figure 4.

**Table 1.** Results tested using two fitting methods on the PGRF dataset, with the best result highlighted in red.

| Model | Index | 100 | 200 |
|---|---|---|---|
| $y = at$ | PSNR/SSIM | 41.22/0.96 | 38.51/0.94 |
| $y = a\sqrt{t}$ | PSNR/SSIM | 41.81/0.97 | 39.50/0.95 |

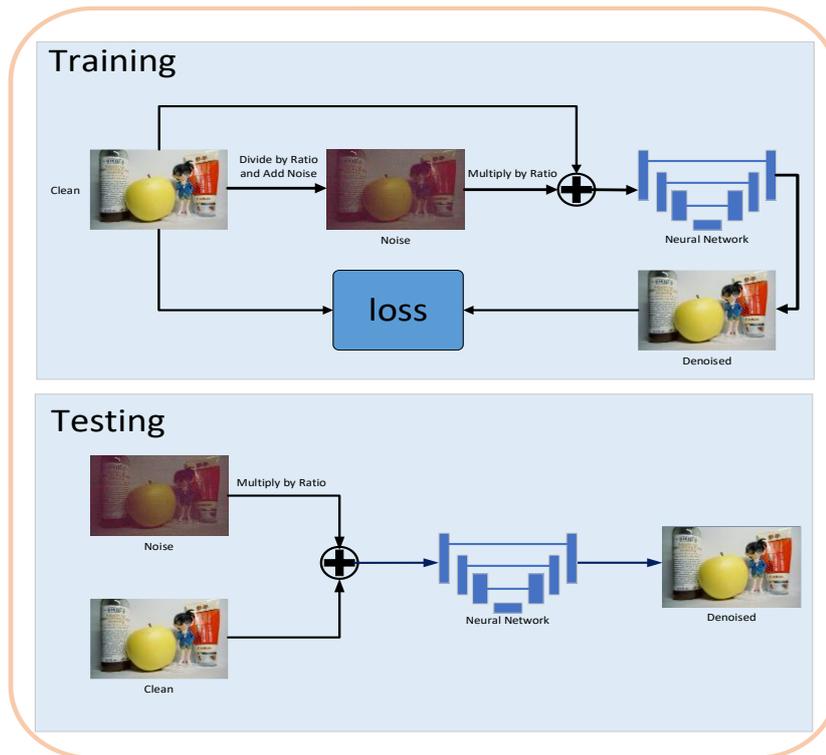

**Figure 4.** Training and Testing Pipeline for Images

5.2.2 Experimental Results

Table 2 summarizes the denoising performance under different exposure ratios on the LLID dataset across various noise models, while Figure 5 presents the visual comparison on the same dataset. "SPC" denotes the results obtained by training the network with our proposed noise generation strategy, where we benchmark against the official pre-trained models released by each method.

For physics-based noise models, the synthetic data of P-G [23] exhibits substantial discrepancies from real sensor noise, leading to noticeable color distortions in the denoised results. Although ELD [7] improves the modeling of signal-independent noise, it lacks the ability to capture fixed pattern noise and dark current shot noise, resulting in suboptimal restoration performance.

For real-data-based models, their applicability is restricted by the labor-intensive process of dataset collection. Moreover, due to the inherent complexity of real-world noise, [18] faces challenges in accurately learning the real noise distribution. While [20] enhances the learning capability of neural networks, the restoration results remain unsatisfactory. In contrast, our approach benefits from precise noise modeling, yielding superior image recovery quality.

**Table 2.** Denoising results for RAW images on the LLID dataset. The best and second-best results are highlighted in red and blue, respectively.

| Ratio | Index | Physics-based | | | | Real noise-based | | |
|---|---|---|---|---|---|---|---|---|
| | | PGRF | SPC | ELD[7] | P-G[23] | LRD[18] | PMN[9] | Paired Data[3] |
| 100 | PSNR | 42.93 | 41.81 | 40.36 | 40.61 | 40.57 | 42.01 | 40.25 |
| | SSIM | 0.97 | 0.97 | 0.96 | 0.94 | 0.97 | 0.97 | 0.96 |
| 200 | PSNR | 40.84 | 39.50 | 38.65 | 37.27 | 39.04 | 39.41 | 38.51 |
| | SSIM | 0.96 | 0.95 | 0.95 | 0.88 | 0.96 | 0.95 | 0.95 |

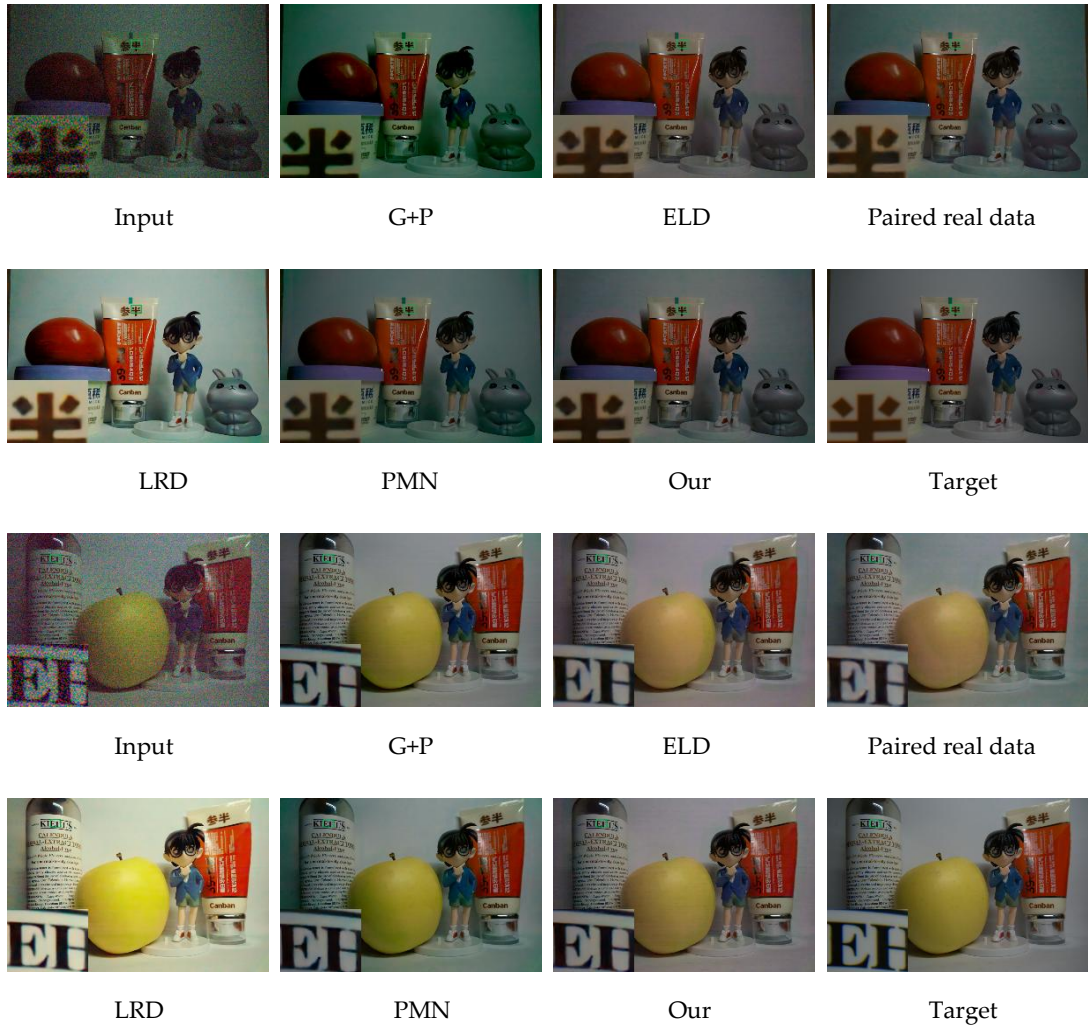

**Figure 5.** We perform a qualitative comparison between our method and competing approaches on the PGRF test set. The regions of interest are highlighted with green boxes for closer inspection upon zooming in.

**5.**2.3 Ablation Study

To validate the effectiveness of our proposed method, we also conducted comparative experiments. Specifically, P denotes photon shot noise, G denotes source follower noise, H denotes row noise, Q denotes quantization noise, F denotes fixed-pattern noise, A denotes dark current shot noise, and noD denotes that the single-pixel calibration method is not used. The evaluation results are summarized in Table 3.

**Table 3.** Ablation study results comparison.

| Model | Index | 100 | 200 |
|---|---|---|---|
| P+G | PSNR/SSIM | 40.64/0.94 | 37.23/0.87 |
| P+G+F | PSNR/SSIM | 40.88/0.97 | 38.46/0.95 |
| P+G+F+H | PSNR/SSIM | 40.93/0.97 | 38.48/0.95 |
| P+G+F+H+Q | PSNR/SSIM | 41.15/0.97 | 38.57/0.95 |
| P+G+F+H+Q+onD | PSNR/SSIM | 40.81/0.96 | 38.44/0.95 |
| P+G+F+H+Q+A | PSNR/SSIM | 41.81/0.97 | 39.50/0.96 |

*5.3 PGRF Model Validation*

To further enhance the quality of image restoration in low-light environments, we thoroughly validated the effectiveness of the PGRF framework and conducted extensive ablation experiments to verify the efficacy of our designed modules. The results demonstrate that our proposed framework achieves superior restoration performance and generalization capability.

5.3.1 Experimental Details

We first used long-exposure images from the SID dataset as clean reference images for training. The SR3UNet network architecture was employed to predict the noise map $v_\theta$. We adopted the Adam optimizer with a batch size of 12 and an initial learning rate of $1\times10^{-4}$, training for 450k iterations. For the trained network, we used a sampling search strategy to determine the optimal sampling time $t_2$, where the step size $s$ was set to 0.1. The optimally processed image was subsequently obtained using the derived $t_2$. The corresponding processing pipeline is illustrated in Figure 6.

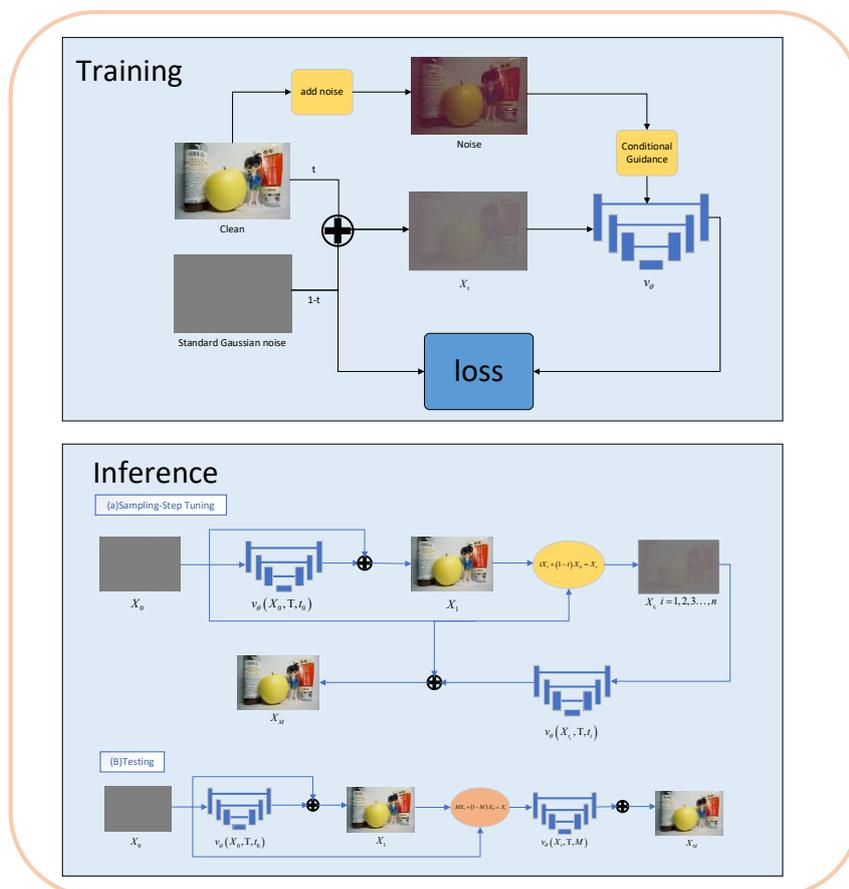

**Figure 6.** Training and Testing Pipeline for Images

5.3.2 Experimental Results

We further compare the denoising performance of the proposed PGRF method against existing approaches on the LLID and ELD datasets, as reported in Tables 2 and 4. Figure 7 illustrates the comparison on the public ELD dataset. The results of the other competing methods, consistent with the analysis in Section 5.2.2, indicate that the PGRF method demonstrates clear superiority in image restoration. For the competing approaches, we directly employed their officially released pretrained models and implementations to obtain the results.

**Table 4.** Denoising results for RAW images on the ELD dataset. The best and second-best results are highlighted in red and blue, respectively.

| Ratio | Index | Physics-based | | | Real noise-based | | |
|---|---|---|---|---|---|---|---|
| | | PGRF | ELD[7] | P-G[23] | LRD[18] | PMN[9] | Paired Data[3] |
| 100 | PSNR | 45.95 | 44.71 | 43.11 | 44.91 | 45.85 | 44.01 |
| | SSIM | 0.97 | 0.97 | 0.92 | 0.98 | 0.98 | 0.97 |
| 200 | PSNR | 43.92 | 42.88 | 39.88 | 42.75 | 43.86 | 42.10 |
| | SSIM | 0.95 | 0.95 | 0.86 | 42.75 | 0.97 | 42.10 |

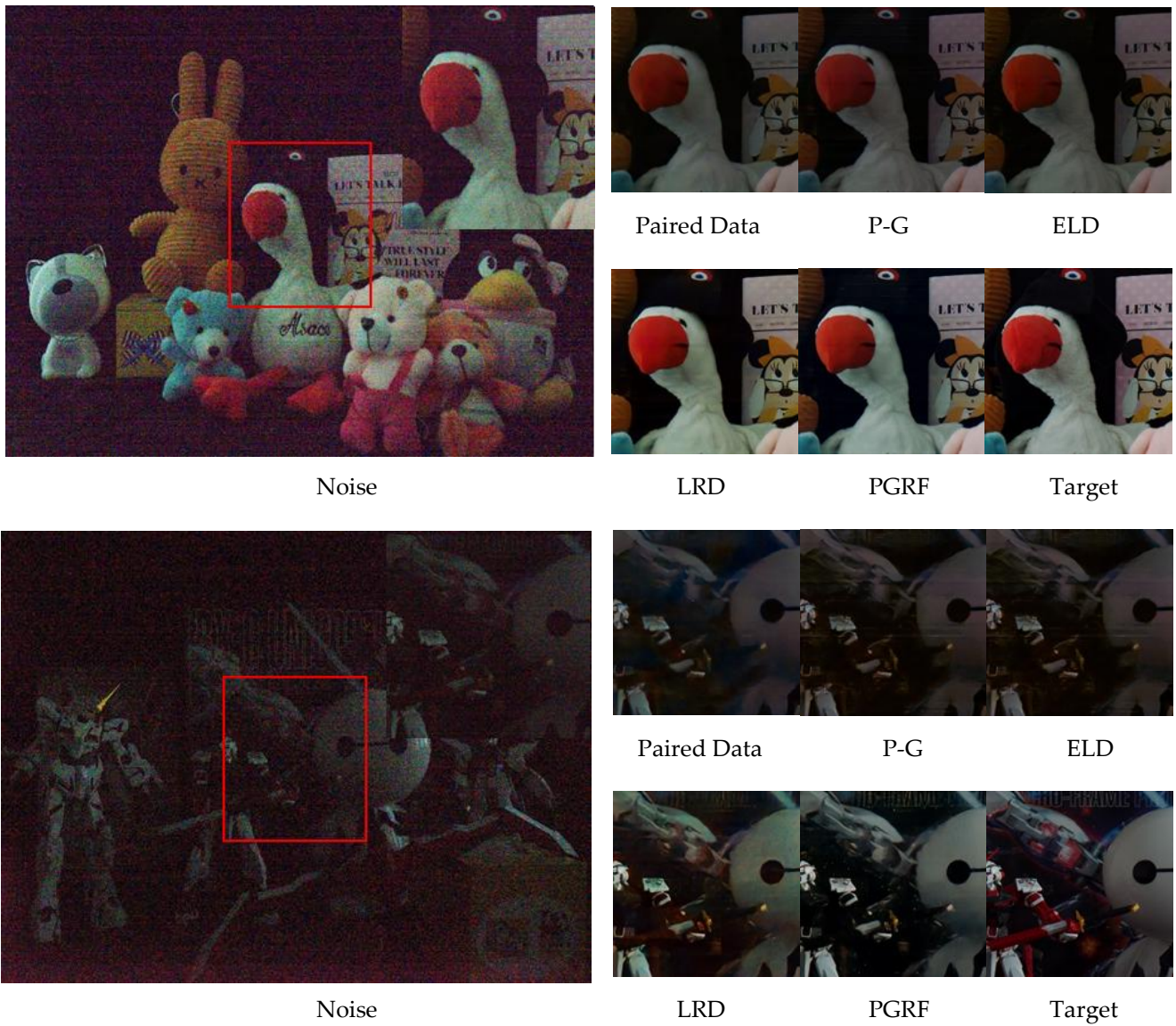

**Figure 7.** We perform a qualitative comparison between our method and competing approaches on the ELD test set. The regions of interest are highlighted with red boxes for closer inspection upon zooming in.

5.3.3 Ablation Study

To further validate the effectiveness of our method, we conducted ablation experiments. In these experiments, a represents results obtained from direct training without using the sampling search strategy, while b represents results with the sampling search strategy applied. As shown in Table 5, the results demonstrate the effectiveness of our sampling search strategy.

Table 5. Ablation study results comparison.

| Model | Index | a | b |
|---|---|---|---|
| 100 | PSNR/SSIM | 45.79/0.97 | 45.85/0.97 |
| 200 | PSNR/SSIM | 43.63/0.95 | 43.68/0.95 |

## 6 Conclusion

This study addresses the challenge of inaccurate noise modeling in synthetic training data for low-light RAW image enhancement. From the perspective of the physical imaging mechanism of image sensors, we propose a composite noise modeling approach that integrates both additive and multiplicative noise. Notably, we extend noise calibration from a global level to the per-pixel level to capture spatially non-uniform noise characteristics caused by microscopic variations in CMOS manufacturing. This per-pixel physical noise simulation significantly improves the consistency between synthetic and real captured data in terms of noise distribution, thereby providing higher-fidelity paired samples for training low-light image enhancement networks.

In terms of generative modeling, this work is the first to combine physical noise modeling with the rectified generative framework Rectified Flow, resulting in the PGRF (Physics-Guided Rectified Flow) framework for low-light image enhancement. By introducing a physics-based conditional control mechanism, the framework effectively constrains the stochasticity of the generation process, achieving a two-step generation from low-light images to target enhanced images. This approach significantly improves generation efficiency while preserving rich detail and natural texture structures.

To validate the proposed method, we constructed the Sony A7S2 indoor low-light dataset (LLID), covering multiple exposure settings and typical indoor scenes. Experimental results demonstrate that PGRF achieves significant improvements in both visual quality and inference efficiency compared with existing methods for low-light RAW image enhancement. This study not only overcomes the limitations of traditional noise modeling but also provides a novel approach for integrating physical priors with efficient generative models, offering important academic value and practical potential.